\documentstyle[aps, pra, graphics, epsfig, floats]{revtex}

\twocolumn

\tighten \draft

\title{Young's experiment and the finiteness of information}

\author{\v Caslav Brukner and Anton Zeilinger}

\address{Institut f\"ur Experimentalphysik, Universit\"at Wien,  Boltzmanngasse 5, A--1090
Wien, Austria}

\date{\today}

\begin{document}

\maketitle

\begin{abstract}

Young's experiment is the quintessential quantum experiment. It is
argued here that quantum interference is a consequence of the
finiteness of information. The observer has the choice whether
that information manifests itself as path information or in the
interference pattern or in both partially to the extent defined by
the finiteness of information.

\end{abstract}


\section{Introduction}

Young's experiment, originally the definitive proof of the wave
nature of light, commands an essential role in the discussion of
the foundations of quantum mechanics. For example, in the
Bohr-Einstein-Dialogue \cite{bohr}, the double-slit experiment was
used as a gedanken experiment with individual quanta. In that
discussion, Einstein wanted to argue that quantum mechanics is
inconsistent in the sense that one can have path information and
observe the interference pattern at the same time, while Bohr was
always able to demonstrate that Einstein's point of view was not
correct. Indeed, if one carefully analyzes any situation where it
is possible to fully know the path the particle took, the
interference pattern cannot be observed. Likewise, if one observes
the full interference pattern, no path information is available.

Young's experiment today is considered the most beautiful
demonstration of the concept of quantum superposition [Fig. 1].
Whenever we do not know, not even in principle, which of the two
paths the particle takes, the quantum state can be written as
\begin{eqnarray}
|\psi\rangle = \frac{1}{\sqrt{2}} &(&|\mbox{passage through left
slit}\rangle \label{particle} \\ &+ &|\mbox{passage through right
slit}\rangle). \nonumber
\end{eqnarray}
In that case, no information whatsoever is available about the
slit the particle passes through. Indeed, if one asked which path
the particle takes in an experiment for a specific run, one would
find the particle in either slit with equal probability.

Yet, obviously, this requires the use of detectors. If one places
one detector each into each slit and if one describes the detector
states by quantum mechanics, then, clearly, the quantum state of
the whole system becomes
\begin{eqnarray}
|\psi\rangle = \frac{1}{\sqrt{2}} &(& |D_L\rangle |\mbox{passage
through left slit}\rangle \label{detectors} \\ &+&|D_R\rangle
|\mbox{passage through right slit}\rangle) \nonumber.
\end{eqnarray}
The description of not only the particle considered, but also the
detector by a quantum state as given in (\ref{detectors}) only has
the meaning that the property of the particle to take a definite
path is related to a property of the detectors. The two detector
states $|D_L\rangle$ and $|D_R\rangle$ describe the detector
having registered the particle passing through the left and right
slit, respectively. These can even be states of an internal degree
of freedom of the interfering particle (e.g. spin or polarization
states or internal atomic states).

\begin{figure}
\centerline{\psfig{width=8.0cm,file=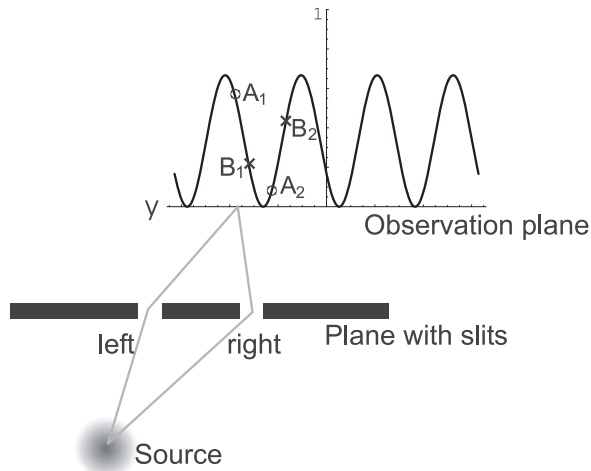}} \caption{
Principle of Young's experiment. The four points $A_1$, $B_1$,
$A_2$ and $B_2$ are chosen such as to determine the information
content in the interference fringes (see text).} \label{fig1}
\end{figure}

A proposal for such an experiment has been made by Scully {\it et
al.} \cite{scully}. Summhammer {\it et al.} \cite{summhammer}
performed a neutron interference experiment and D\"{u}rr {\it et al.}
\cite{duerr} performed an atomic interference experiment where the
disappearance of the interference pattern has to be attributed to
the correlations between the internal neutron or atomic states,
which serve as which-path detectors, and the paths taken inside
the interferometer. In these experiments the loss of interference
is due to the fact that path information is available, in
principle, independent of the fact whether the experimentalist
cares to read it out or not.

If the two detector states are orthogonal, then the two particle
states cannot interfere, as Eq. (\ref{detectors}) describes then a
maximally entangled state and thus one could determine the path of
the particle by observing the detector state. Only if the two
detector states are not orthogonal \cite{wootters} or if they are
projected by a measurement onto a state that is orthogonal to
neither one of them \cite{scully,scully+} then path interference
of a certain contrast may reappear, as then the complete knowledge
about the path is not available.

\section{Coherence and path information in interference experiment with
fullerenes}

Technological progress in the times since the
Bohr-Einstein-Dialogue made it possible to realize quantum
interference with many different particles all the way to massive
molecules, like the fullerenes \cite{arndt,nairz,arndt+} C-60 and
C-70. It is interesting to note that in the latter experiment, the
fullerene molecules are at temperatures as high as 900 K. This
implies that they are not completely decoupled from the
environment. On the contrary, they typically emit a few photons on
their path from the source to the detector \cite{arndt+}. So why
do interference fringes still appear [Fig. 2]? Could one not use
the emitted photons to trace the path of the fullerene? The reason
can easily be understood by referring to Eq. (\ref{detectors}).
The wavelength of the emitted photons is typically of the order of
a few micrometers, which has to be compared to the path
separation, which is much lower. Therefore, the states of the two
photons emitted by a fullerene on either of the interfering paths
are nearly identical, implying that the photons carry virtually no
information into the environment.

The modulus of the scalar product between the two states of the
photons corresponding to the emission by a fullerene on either of
the interfering paths can be used to quantify the information
about the path of the fullerene, which can in principle be
extracted if the photons were observed. Only if the scalar product
is non-zero, then an interference pattern of a certain contrast
may appear, as then the path is not completely known. In general,
the contrast (visibility $V$) of the interference pattern is equal
to the modulus of the scalar product between the two detectors
states, $V=|\langle D_R|D_L \rangle|$. We now calculate the scalar
product between the two photon states which serve as detector
states in the fullerene experiment.

For the reason of simplicity we consider the fullerene experiment
as a double-slit experiment. Suppose that the interfering
fullerene emits $N$ photons at the moment it reaches the screen
with the two slits. That is, the photons are emitted by the
fullerene either at the left slit or at the right slit. Then the
visibility $V$ of the fullerene interference pattern at the
observation screen is equal to the modulus of the following scalar
product
\begin{equation}
V\!=\!| \langle \mbox{N photons from left slit}| \mbox{N photons
from right slit} \rangle|. \label{vuk}
\end{equation}
Because the two possible states are the same for every of the N
photons, one can transform Eq. (\ref{vuk}) into
\begin{equation}
V = \left|\int d\vec{r} \phi(\vec{r},\vec{r}_L)
\phi^*(\vec{r},\vec{r}_R)\right|^N, \label{srce}
\end{equation}
where
\begin{equation}
\phi(\vec{r},\vec{r}_L) =
\frac{e^{iK|\vec{r}-\vec{r}_L|}}{|\vec{r}-\vec{r}_L|} \mbox{ and }
\phi(\vec{r},\vec{r}_R) =
\frac{e^{iK|\vec{r}-\vec{r}_R|}}{|\vec{r}-\vec{r}_R|}
\end{equation}
are the two amplitudes (spherical waves) of a photon at
observation point $\vec{r}$, which are emitted from the point
source localized at the position $\vec{r}_L$ of the left slit and
$\vec{r}_R$ of the right slit, respectively. Here $K$ is the
wave-number of the photon.

To calculate the integral in Eq. (\ref{srce}) we use the
substitution $\xi f=\frac{|\vec{r} - \vec{r}_R| + |\vec{r} -
\vec{r}_L|}{2}$ and $\eta f =\frac{|\vec{r} - \vec{r}_R| -
|\vec{r} - \vec{r}_L|}{2}$ and perform an integration over prolate
spheroidal coordinates within the intervals: $1\! \leq \!\xi\! <\!
\infty$, $-1 \! \leq \! \eta\! \leq \! 1$ and $0 \!\leq\! \phi
\!\leq \!2\pi$. The integration volume is $d\vec{r} = d\eta d\xi
d\phi |f^3(\eta^2-\xi^2)|$. Using straightforward algebra one
obtains
\begin{equation}
V\propto \left|\frac{\sin(Kd)}{Kd}\right|^N, \label{mrmor}
\end{equation}
where $2f=d$ and $d$ is the separation between two slits.

Such dependence of the visibility on the number $N$ of emitted
photons and their wave-number $K$ is in agreement with decoherence
observed in an atom interferometry \cite{pritchard}. It is now
clear from Eq. (\ref{mrmor}) that in the extreme case of the wave
length much smaller then the slit separation and/or sufficiently
large number of emitted photons the visibility $V$ vanishes. Yet,
in the fullerene experiment another extreme case is reached. There
the slit separation $d=1\mu m$, the photons wave length is of the
order of $10\mu m$, and the estimated number of photons emitted
during the entire time of flight of the fullerene are 1-2.
Therefore $\left|\frac{\sin(Kd)}{Kd}\right|^N \approx 1 $ and the
high visibility remains preserved.

\begin{figure}
\centerline{\psfig{width=9.0cm,file=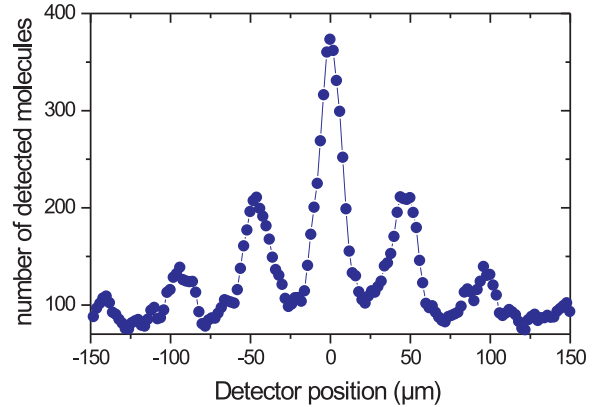}}
\caption{Interference pattern of C-60 molecules behind a 100 nm
grating, which proves the absence of thermal decoherence in the
experiment [6], even for molecules with internal temperatures as
high as 900 K (M. Arndt, O. Nairz, and A. Zeilinger, in
preparation).} \label{fig3}
\end{figure}

\section{Information and complementarity in a quantum interference experiment}

The possible choice between path information and the observability
of interference patterns is one of the most basic manifestations
of quantum complementarity, as introduced by Niels Bohr. Following
our discussion, it is clear that it is the experimentalist who
decides which observable to measure. He can decide, for example,
whether to put a detector into the interfering paths or not. This
role of the observer has led to numerous misunderstandings about
the Copenhagen interpretation of quantum mechanics. Very often,
and erroneously, a strong subjective element is brought into the
discussion, implying that even the consciousness of the observer
has a role in the quantum measurement process. One has to be very
careful at this point.

Just to follow our example, the observer can decide whether or not
to put detectors into the interfering path. That way, by deciding
whether or not to determine the path through the two-slit
experiment, he can decide which property can become reality. If he
chooses not to put the detectors there, then the interference
pattern will become reality; if he does put the detectors there,
then the beam path will become reality. Yet, most importantly, the
observer has no influence on the specific element of the world
which becomes reality. Specifically, if he chooses to determine
the path, he has no influence whatsoever which of the two paths,
the left one or the right one, Nature will tell him is the one
where the particle is found. Likewise, if he chooses to observe
the interference pattern he has no influence whatsoever where in
the observation plane he will observe a specific particle. Both
outcomes are completely random.

We therefore argue that the observer has a qualitative influence
on Nature by deciding via his choice of apparatus which quality
can manifest itself as reality, but he has no quantitative
influence in the sense of which specific result will be the
outcome. It therefore appears that the objective randomness of
quantum measurement provides a limit to the control any
experimentalist has. Bohr \cite{bohr1934} writes succinctly: ''...
a subsequent measurement to a certain degree deprives the
information given by a previous measurement of its significance
for predicting the future course of phenomena. Obviously, these
facts not only set a limit to the extent of the information
obtainable by measurement, but they also set a limit to the
meaning which we may attribute to such information. We meet here
in a new light the old truth that in our description of nature the
purpose is not to disclose the real essence of the phenomena but
only to track down, so far as it is possible, relations between
the manifold aspects of our experience.''

We will now argue that the impossibility of joint perfect
observation of both path and the interference pattern is a natural
consequence of the finiteness of the information content of a
quantum system. On the basis of a specific measure of information
we will define information content of a quantum system. That
information can fully be contained either in the path or in the
interference pattern. In both of them only partially to the extent
defined by the fundamental limit on the information content.
Therefore we will give a quantitative information-theoretic
formulation of quantum complementarity in Young's experiment.

In a double-slit experiment the path information is a dichotomic,
i.e. a two-valued observable while the position in the
interference pattern is a continuous one, which makes the
consideration more complicated.  For that reason we will modify
our set-up to that of an interferometer [Fig. 3] where both path
information and interference observation are dichotomic.
Afterwards we will extend our analysis to a double-slit
experiment. If in Fig. 3 the incoming state $\psi_1$ has amplitude
$a$ and the incoming state $\psi_2$ has amplitude $b$ ($a,b \in
{\bf R}, a^2+b^2=1)$, then by the usual rules of a symmetric beam
splitter \cite{zeilinger}, the outgoing states $\psi_3$ and
$\psi_4$ become
\begin{equation}
\psi_3=\frac{1}{\sqrt{2}} (iae^{i\chi}+b) \hspace{0.3cm}
\psi_4=\frac{1}{\sqrt{2}} (ae^{i\chi}+ib),
\end{equation}
where we allow for an arbitrary, but constant, phase difference
$\chi$ between amplitudes $a$ and $b$. It now follows that the
probabilities $p_1$, $p_2$, $p_3$, and $p_4$ to find an individual
particle in any of the four beams are:
\begin{eqnarray}
p_1&=&a^2, \hspace{0.3cm} p_2=b^2 \nonumber \\
p_3&=&\frac{1}{2}(1-2ab\sin\chi),  \hspace{0.3cm}
p_4=\frac{1}{2}(1+2ab\sin\chi).
\end{eqnarray}
Evidently, because of unitarity, $p_1+p_2=1$ and $p_3+p_4=1$. How
can we see now the complementarity between the path information
and the interference phenomenon?

\begin{figure}
\centerline{\psfig{width=6.7cm,file=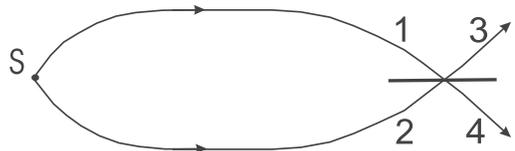}}
\caption{A quantum interferometer. The source 'S' emits coherent
spherical waves, of which two rays are selected and are incident
on the beam splitter. Each of the two rays then has the same
amplitude for being transmitted or reflected at the beam-splitter,
and thus the outgoing beams 3 and 4 are coherent superpositions of
the incoming ones 1 and 2.} \label{fig2}
\end{figure}

It is suggestive to assume that our ability to determine which
path the particle takes is related to the modulus $|p_1-p_2|$ of
the difference between the probabilities in path 1 and path 2.
This difference results in the minimal value of 0 if both
probabilities are equal and in the maximal value of unity if one
of the probabilities is 1. In the same way as we assume the
information available about the path to be proportional to the
modulus of the difference $|p_1-p_2|$, we may also assume the
information in the interference pattern to be proportional to the
modulus of the difference $|p_3-p_4|$. There is some
complementarity between $|p_1-p_2|$ and  $|p_3-p_4|$, and we will
now express it quantitatively such that the total information is a
constant. Indeed, we find, if we introduce our new measure of
information \cite{brukner} we are led to a quantitative statement
of the complementarity principle. Our new measure of information,
which is suitable to define the information gain in a quantum
experiment, takes probability squares as a quantitative statement
of our knowledge. In \cite{brukner} it was shown that this
particular measure of information is related to the estimation of
the future number of occurrence of a specific outcome in a
repetition of a binary experiment with two probabilistic outcomes.

We now introduce the following quantitative amounts of information
\begin{equation}
I_1\!=\!(p_1-p_2)^2, \mbox{ } I_2\!=\!(p_3-p_4)^2, \mbox{ and }
I_3\!=\!(p'_3-p'_4)^2, \label{measures}
\end{equation}
where we have introduced the probabilities $p'_3$ and $p'_4$ as
those probabilities where we use an additional phase shifter of
phase $\frac{\pi}{2}$ in, say, beam 2, resulting in the
probabilities
\begin{equation}
p'_3=\frac{1}{2}(1-2ab\cos\chi), \mbox{ and }
p'_4=\frac{1}{2}(1+2ab\cos\chi).
\end{equation}
The reason that we consider also the probabilities $p'_3$ and
$p'_4$ is that for any specific phase shifts $\chi$ between the
two incoming amplitudes, even without path information, our
knowledge whether the particle will be found in beam 3 or 4 might
not be maximal (Fig. \ref{fig2}). This knowledge however can then
be re-established if an additional phase shift of $\frac{\pi}{2}$
is introduced between the two amplitudes.

Now, for the sum of the three individual measures of information,
we obtain
\begin{equation}
I_1+I_2+I_3=1. \label{content}
\end{equation}
Such a complementarity relation resulting in a constant is
possible only if our new measure (\ref{measures}) is used and
could not be obtained if, for example, Shannon's measure of
information were used \cite{brukner+}. An important property of
the information content of a quantum system as defined by Eq.
(\ref{content}) is that it neither depends on the incoming
amplitudes $a$ and $b$, nor on the phase factor $\chi$ between
them. This means that the total information is invariant under
unitary transformations and thus equal for all possible pure
incoming states. Therefore different pure incoming states might
have different individual measures of information $I_1$, $I_2$ and
$I_3$ but their sum is always 1 bit of information.

Here $I_1$  describes the path information and $I_2$ and $I_3$
together describe the visibility of the interference effect. We
may therefore introduce the new variables $I_{path}=I_1$ and
$I_{interf}=I_2+I_3$, and we obtain the final result (See also
\cite{englert}.)
\begin{equation}
I_{path} + I_{interf} =1. \label{central}
\end{equation}
which is a quantitative statement of the principle of
complementarity in Young's experiment. One may reinterpret Eq.
(\ref{central}) such that a single particle in Young's experiment
is just the representative of one bit of information and the
experimentalist has the choice by deciding whether to determine
the path or not, whether this information resides in the path or
in interference or in both of them partially to the extend defined
by Eq. (\ref{central}).

We will now extend our consideration to the situation of a
double-slit experiment [Fig. 1]. We assume that the amplitude of
the interfering particle is $a$ in the left slit and $b$ in the
right slit $(a,b \in R, a^2+b^2=1)$, where again we allow for an
arbitrary phase difference $\chi$ between the two amplitudes. A
typical interference pattern in the Fraunhofer limit has a
sinusoidal form with a periodicity of $Y=\frac{2 \pi L}{kd}$ where
$k$ is the de-Broglie wave-number, $d$ is the separation between
the two slits and  $L$ is the distance between the plane with
slits and the observation plane.

Consider now two pairs of points $A_1=y$, $A_2=y+Y/2$ and
$B_1=y+Y/4, B_2=y+3Y/4$ in the observation plane, as shown in Fig.
1. On the basis of our new measure of information we now introduce
the amount of information
$I_A=\left(\frac{p(A_1)}{p(A_1)+p(A_2)}-\frac{p(A_2)}{p(A_1)+p(A_2)}\right)^2$
for the pairs of points $A_1$ and $A_2$, and similarly
$I_B=\left(\frac{p(B_1)}{p(B_1)+p(B_2)}-\frac{p(B_2)}{p(B_1)+p(B_2)}\right)^2$
for $B_1$ and $B_2$. Here, for example,
$\frac{p(A_1)}{p(A_1)+p(A_2)}$ is the conditional probability to
detect particle at $A_1$ given that the particle is to be found
either in $A_1$ or $A_2$. Therefore $I_A$ is the measure of the
information that the particle will be found in the specific point
$A_1$ or in the specific point $A_2$ given that we know it will be
found at $A_1$ or $A_2$ anyway. The probability density to detect
the particle at point $y$ in the observation plane in the
Fraunhofer limit is given by
\begin{equation}
p(y)= \frac{1}{Y}\left[ 1+2ab
\cos\left(\frac{kd}{L}y+\chi\right)\right].
\end{equation}
Here the probability distribution is normalized such that the
total probability to find the particle somewhere within the
interval $[0,Y]$ of one period is unity. If we now use $I_1$ for
the amount of information contained in the path and $I_A$ in the
pair of observation points $A_1$, $A_2$ and $I_B$ in the pair
$B_1$, $B_2$, then we obtain again that $I_1+I_A+I_B=1$.

We notice that the four selected points $A_1$, $A_2$, $B_1$ and
$B_2$ for which the probability is calculated are just separated
by $Y/4$ and can be selected for any choice of $y$. Like in the
case of the interferometer, we will now summarize all individual
measures of information $I_A$ and $I_B$ for all $y$ and thus
obtain the information contained in the full interference pattern.

We still use $I_{path}$ as given above for the measure of
information contained in the path. Yet now we suggest the
information contained in the interfering path to be defined by the
integral
\begin{equation}
I_{interf}= 2Y \int_{0}^{Y/2}[p(y)-p(y+Y/2)]^2 dy \label{integral}
\end{equation}
Note that the integrand in Eq. (\ref{integral}) contains the
combinations
\[
[p(y)-p(y+Y/2)]^2 +[p(y+Y/4)-p(y+3Y/4)]^2
\]
for every $y$ within the interval $[0,Y/4)$, which correspond
exactly to the sum $I_A+I_B$ introduced above. One can easily
calculate that $I_{interf}=4a^2b^2$. Therefore we have again
$I_{path}+I_{interf}=1$ for the sum of the measures of information
contained in the path and in the interference pattern.

The discussion presented above obviously is just one specific
example of quantum complementarity at work. It is obvious that
this can be extended to much more complicated situations, as for
example to the notion of quantum entanglement \cite{brukner++}.
From a fundamental perspective, this approach suggests that the
most basic notion of quantum mechanics is information
\cite{zeilinger+}.

This work was supported by the Austrian Science Foundation, SFB
project No. S6505 and by project ERB-FMRX-CT-96-0087 of the
European Commission.

\end{document}